# Multiresolution Priority Queues and Applications

Jordi Ros-Giralt, Alan Commike, Peter Cullen Richard Lethin
{giralt, commike, cullen, lethin}@reservoir.com
Reservoir Labs
632 Broadway Suite 803, New York, NY 10012

*Abstract* − **Priority queues are container data structures essential to many high performance computing applications. In this paper, we introduce *multiresolution priority queues*, a data structure that achieves greater performance than the standard heap based implementation by trading off a controllable amount of resolution in the space of priorities. The new data structure can reduce the worst case performance of inserting an element from $O(log(n))$ to $O(log(r))$, where $n$ is the number of elements in the queue and $r$ is the number of resolution groups in the priority space. The worst case cost of extracting the top element is $O(1)$. When the number of elements in the table is high, the amortized cost to insert an element becomes $O(1)$.**

## 1. Introduction

Priority queues are a type of *container data structures* designed to guarantee that the element at the front of the queue is the greatest of all the elements it contains, according to some total ordering defined by their *priority*. This allows for elements of higher priority to be extracted first, independently of the order in which they are inserted into the queue. Fast implementations of priority queues are important to many high performance computing (HPC) applications and hence optimizing their performance has been the subject of intense research. In graph searching, the algorithmic complexity of essential algorithms such as finding the shortest path between two nodes in a graph rely entirely on the performance of a priority queue (see for instance Dijkstra's algorithm [1], [2]). Other essential applications are found in the field of bandwidth management. For instance, in the Internet, routers often have insufficient capacity to transmit all packets at line rate and so they use priority queues in order to assign higher preference to those packets that are most critical in preserving the quality of experience (QoE) of the network (e.g., packets carrying delay sensitive data such as voice). Priority queues are also important in other areas of research including Huffman compression codes, operating systems, Bayesian spam filtering, discrete optimization, simulation of colliding particles, or artificial intelligence to name some examples.

The work we present stems from research on improving the performance of a priority queue by introducing a trade-off to exploit an existing gap between the applications' requirements and the set of priorities supported by a priority queue. With this objective, we introduce the concept of *multiresolution priority queue*, a new container data structure that can flexibly trade off a controllable amount of resolution in the space of priorities in exchange for achieving greater performance.

A simple example works as follows. Suppose that we use a priority queue to hold jobs $\{job_1, job_2, ..., job_n\}$ that need to be scheduled using a "shortest job first" policy. Each job $job_i$ takes a time to process equal to $p_i$, and the removal of an element from the queue returns the job with the smallest processing time $min\{p_i\}$. (Hence $p_i$ determines the priority of each job.) A common approach is to use a binary heap based implementation of a priority queue, which effectively provides support for all possible range of priorities $p_i$ in the space of real numbers $\mathbb{R}$. Assume now that the set of possible priority values $p_i$ is discretized into $l$ subsets and that the application only requires guaranteeing the ordering between elements belonging to different subsets. A multiresolution priority queue is a data structure that can provide better performance than a standard priority queue by supporting the coarser grained resolution space consisting of $l$ subsets rather than the complete set of priorities.

Multiresolution priority queues can also be understood as a generalization of standard priority queues in which the total ordering requirement imposed on the set of priorities is relaxed, allowing for elements in the queue to be only partially ordered. Hence an alternative name to refer to this type of container data structures would be *partial order priority queues*.

This paper is organized as follows. In Section 2, we summarize some of the main historical results in the research area of high performance priority queues. Section 3 presents the body of our work, introducing a formal definition of the concept of multiresolution

---

This work was funded in part by the US Department of Energy under contract DE-SC0017184.
A patent application with Serial No. 62/512,438 has been submitted for this work.

priority queues, providing pseudocode for the main routines to *insert* and *extract* elements from the queue, and analyzing the algorithm's complexity. This section introduces also two variants of the base algorithm to support *sliding priority windows* for applications such as event-driven simulators and a *binary tree* based optimization applicable to general graph algorithms such as finding the shortest path between two nodes. Section 4 describes our implementation of a multiresolution priority queue and demonstrates how it can help a real world HPC application achieve greater performance than a traditional binary heap based solution. We summarize the main conclusions of this work in Section 5.

## 2. Background Work

Priority queues have been the subject of intense research since J. W. J. Williams introduced the *binary heap* data structure and the *heapsort* algorithm in 1964 [3]. Williams' work lead to a very efficient implementation with a worst case cost to insert and extract elements from the priority queue of $O(log(n))$, where $n$ is the number of elements in the queue. This is in many applications a preferred way to implement priority queues due to its simplicity and good performance. Variations exist that can achieve better performance on some operations at the expense of adding more implementation complexity, such as Fibonacci heaps [4], which have an amortized cost of inserting and removing an element of $O(1)$ and $O(log(n))$, respectively.

*Soft heaps* by Chazelle [5] relate to our work in that they also achieve a constant amortized cost by introducing a certain amount of error. Specifically, a soft heap can break the information-theoretic barriers of the problem by introducing error in the key space in a way that the problem's entropy is reduced. While soft heaps provide the most asymptotically efficient known algorithm for finding minimum spanning trees [6], their applications are limited in that they cannot guarantee an upper bound error on the number of elements in the queue. (Depending on the sequence of insertions and extractions, all keys could be corrupted.) Instead, following on Chazelle's analogy, a multiresolution priority queue breaks the information-theoretic barriers of the problem by exploiting the multiresolution properties of its priority space. Since in many problems the entropy of the priority space is lower than the entropy of the key space, the end result is a container data structure with a lower computational complexity. In particular, a multi-resolutive priority space will have a constant entropy independent of the number of elements in the queue, and this result leads to our $O(1)$ bound.

Our approach is perhaps closest to the data structures known as *bucket queues* [7] and *calendar queues* [8]. When the set of priorities are integers and bounded to $C$, a bucket queue can insert and extract elements in $O(1)$ and $O(C)$, respectively. The resolution of this data structure however cannot be tuned and it does not support the general case of priorities in the set of real numbers $\mathbb{R}$. Calendar queues are $O(1)$ for both insert and remove operations, but to achieve these bounds they need to be balanced and they also do not support changing the resolution of the priority set to optimize the trade off performance versus accuracy to match the application requirements.

Our data structure can also be understood as a type of *sketching data structure*. These are data structures that can answer certain questions about the data they store extremely efficiently, at the price of an occasional error. Examples of sketching data structures are Bloom filters [9], count-min sketch [10] and the LFN data structure [11].

## 3. Multi-Resolution Priority Queues

### 3.1. Definition

A multiresolution priority queue (MR-PQ) is a container data structure that retrieves its elements based on their priority and according to three parameters: a *priority resolution* $p_\Delta$ and two *priority limits* $p_{min}$ and $p_{max}$. The priority limits are chosen such that $p_{max} > p_{min}$ and $(p_{max} - p_{min})$ is divisible by $p_\Delta$. Unlike traditional priority queues, multiresolution priority queues do not guarantee the total order of elements within a certain priority resolution $p_\Delta$. However, by trading off a small resolution, they can achieve higher performance. In particular, because many real world applications do not require priority queues to support infinite resolution spaces, in many of these cases MR-PQs can be used to achieve greater performance at no cost, as we will show.

More formally, an MR-PQ is a queue that at all times maintains the following invariant:

*Property 1. Multiresolution Priority Queue (MR-PQ) Invariant.* Let $e_i$ and $e_j$ be two arbitrary elements with priorities $p_i$ and $p_j$, respectively, where $p_{min} \leq p_i < p_{max}$ and $p_{min} \leq p_j < p_{max}$. Then for all possible states, a multiresolution priority queue ensures that element $e_i$ is dequeued before element $e_j$ if the following condition is true:

$$\lfloor (p_i - p_{min})/p_\Delta \rfloor < \lfloor (p_j - p_{min})/p_\Delta \rfloor \qquad (1)$$

§

Intuitively, an MR-PQ discretizes the priority space into a sequence of slots or *resolution groups* $[p_{min}, p_{min} + p_\Delta)$, $[p_{min} + p_\Delta, p_{min} + 2 \cdot p_\Delta)$, ..., $[p_{max} - p_\Delta, p_{max})$ and prioritizes elements according to the slot in which they belong. Those elements belonging to lower slots are given higher priority, while those in the same slot are given the same priority. Within a slot, a multiresolution priority queue does not guarantee the priority ordering of its elements, which enables a mechanism to control the trade off between the resolution of the priority set and the performance of the queue. As we will see, the larger the resolution parameter $p_\Delta$, the higher the performance of the queue, at the expense of making the priority space more granular. In the limit where the priority resolution becomes infinitely small, $p_\Delta \to 0$, an MR-PQ degenerates to a standard priority queue in which the invariant in Property 1 is satisfied for arbitrarily small priority slots, hence achieving infinite resolution.

Notice that in our chosen notation and without loss of generality, elements with smaller values of priority are treated as higher priority. This is also commonly known in the literature as a *min-priority queue* [2]. While this choice is arbitrary and purely conventional, it will make the generalization to support *sliding priorities* in Section 3.4 easier. Well-known examples of applications using *min-priority queues* include event-driven simulators (which will be the subject of our performance benchmarks in Section 4) or shortest path algorithms such as Dijkstra's algorithm [1], among many others. (See also Section 3.5.)

### 3.2. The Base Algorithm

A multiresolution priority queue (MR-PQ) consists of two data structures: a *queue* (`queue`) and a *queue lookup table* (`qlt`). The queue is implemented as a doubly linked list of elements, while the lookup table is implemented as a one dimensional array of pointers to elements. Elements stored in the doubly linked list have a next pointer (`next`) and a previous pointer (`prev`). They include also a third field which is a number representing their priority (`prio`).

Let `pmin`, `pmax`, and `pdelta` be three parameters initialized to $p_{min}$, $p_{max}$ and $p_\Delta$, respectively. Then the following instructions are used to build an empty MR-PQ:

**BUILD(q)**

```
1    sentinel = alloc_element();
2    queue = sentinel;
3    queue->next = queue;
4    queue->prev = queue;
5    qltsize = (pmax-pmin)/pdelta + 1;
6    for i in [1, qltsize):
7        qlt[i] = NULL;
8    qlt[0] = queue;
9    queue->prio = pmin - pdelta;
```

The `queue` object is implemented as an element acting as its sentinel (lines 1 and 2). Lines 3 and 4 set the queue to empty. The lookup table `qlt` is implemented as a vector of pointers to elements in the queue with as many entries as the number of resolution slots $(p_{max} - p_{min})/p_\Delta$ plus one to accommodate for the queue's sentinel. In its initial state, all of its elements are set to `NULL` (since the queue is empty) except for the first element, which is permanently set to the sentinel (lines 6, 7 and 8). In line 9, the sentinel is given a fictitious priority of $p_{min} - p_\Delta$, to ensure that no other element can have a lower priority value than itself. Serving the role of guarding the queue, the sentinel is never removed.

To insert an element `e`, the base MR-PQ algorithm executes the following instructions:

**INSERT(e)**

```
10   slot = slot_iter = QLTSLOT(e);
11   while qlt[slot_iter] == NULL:
12       slot_iter--;
13   if slot_iter == slot: // Add to the left
14       e->next = qlt[slot];
15       e->prev = qlt[slot]->prev;
16       qlt[slot]->prev->next = e;
17       qlt[slot]->prev = e;
18   else:  // Add to the right
19       e->next = qlt[slot_iter]->next;
20       e->prev = qlt[slot_iter];
21       qlt[slot_iter]->next->prev = e;
22       qlt[slot_iter]->next = e;
23   qlt[slot] = e;
```

The invocation of `QLTSLOT(e)` in line 10 calculates an index to the *qlt* table associated with element `e` as follows:

**QLTSLOT(e)**

```
24   slot = (int)((e->prio - queue->prio)/pdelta);
25   return slot;
```

After computing the slot of element `e` in line 10, we perform a search on the `qlt` table for a non-empty slot `slot_iter` starting from index `slot` and decrementing the index at each iteration (lines 11 and 12). Notice that because `qlt[0]` is permanently set to the queue's sentinel, this operation is guaranteed to complete. If the `slot_iter` found is `slot`, it means there is already one or more elements stored in `e`'s slot. In this case, we doubly-link e to the left of the first element found in the slot qlt[slot] (lines 14 through 17). If instead e's slot is empty, then we doubly-link e to the right of the element found in the first non-empty slot qlt[slot_iter] (lines 19 through 22) and set qlt[slot] to point to e (line 23). This ordering strategy is important to ensure that the MR-PQ invariant (Property 1) is preserved at all times, as we will formally demonstrate in Section 3.3. Intuitively, the insertion of an element in the MR-PQ data structure consists in building a list of elements that are doubly linked in a way that it maintains a correct priority ordering according to the resolution parameter $p_\Delta$.

As an example, let a multiresolution priority queue have parameters $p_\Delta = 3$, $p_{min} = 7$ and $p_{max} = 31$, and assume we insert seven elements with priorities 19, 11, 17, 12, 24, 22 and 29 (inserted in this order). Then, Fig. 1 provides the state of the MR-PQ after inserting all the elements. Notice that while elements belonging to different slots (referred in Fig. 1 as resolution groups RG1 through RG5) are guaranteed to be served according to their priority (e.g., the elements with priorities 12 and 11 are served before the element with priority 17), the queue does not guarantee the ordering of elements that belong to the same slot (e.g., the element with priority 12 is served before the element with priority 11).

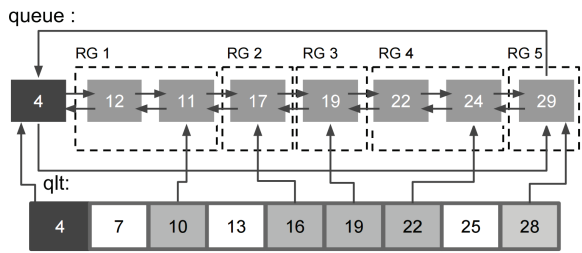

*Fig. 1. The state of a multiresolution priority queue after inserting elements with priorities 19, 11, 17, 12, 24, 22 and 29 (in this order).*

The operation to peek (without removing) the top priority element from the queue consists in retrieving the first element in the queue and checking that it's different than the sentinel:

**PEEK()**

```
26  e = q->next;
27  if e == q:
28      return NULL;  // Queue is empty
29  return e;
```

To extract an arbitrary element from the queue, we unlink it (lines 30 and 31) and invoke QLTREPAIR to repair the lookup table (line 32):

**EXTRACT(e)**

```
30  e->prev->next = e->next;
31  e->next->prev = e->prev;
32  QLTREPAIR(e);
33  return e;
```

If element e is not pointed to by any of the lookup table entries, then there is nothing to repair (lines 35 and 36). Otherwise, if e->prev is in the same slot as e (line 37), then we set qlt[slot] to point to e->prev (line 38) to ensure the slot is pointing to the rightmost element of its resolution group. If instead e->prev is not in the same slot as e, then the slot has become empty and we set qlt[slot] to NULL (line 40).

**QLTREPAIR(e)**

```
34  slot = QLTSLOT(e);
35  if qlt[slot] != e:
36      return; // Nothing to fix
37  if slot == QLTSLOT(e->prev):
38      qlt[slot] = e->prev; // Fix the slot
39  else:
40      qlt[slot] = NULL; // No elements left in slot
```

Now the operation to extract the highest priority element from the queue can be performed with a PEEK() followed by a EXTRACT(e):

**EXTRACTMIN()**

```
41  e = PEEK();
42  EXTRACT(e);
43  return e;
```

*3.3. Correctness and Computational Cost*

In the following two lemmas, we state the correctness and computational cost of the MR-PQ algorithm.

*Lemma 1. Correctness of the MR-PQ algorithm.* The INSERT and REMOVE routines preserve the MR-PQ invariant (Property 1).

*Proof.* Notice first that the PEEK and the EXTRACTMIN routines select elements from the queue by using the sentinel's next pointer (line 26) and therefore to prove the above lemma it suffices to show that equation (1) is preserved for any two consecutive elements $e_i$ and $e_j$ such that $(e_i \rightarrow next) = e_j$. In what follows, we will assume that the queue is in a state in which the MR-PQ invariant is true and demonstrate that after the INSERT or the EXTRACT routines the queue continues to preserve the invariant.

Assume that we invoke the INSERT routine on element $e_i$. There are two cases: $qlt[QltSlot(e_i)]$ is not NULL or $qlt[QltSlot(e_i)]$ is NULL. Assume the first case, then lines 14 through 17 insert element $e_i$ between two elements $e_j$ and $e_k$ such that

$(e_j \rightarrow next) = e_i = (e_k \rightarrow prev)$. After the insertion, the following is true:

- $QltSlot(e_j) = QltSlot(e_i)$ or $QltSlot(e_j) < QltSlot(e_i)$, in which case the invariant is preserved for elements $e_i$ and $e_j$.

- $QltSlot(e_k) = QltSlot(e_i)$, which implies that the invariant is also preserved for elements $e_i$ and $e_k$.

Assume now that $qlt[QltSlot(e_i)]$ is NULL, then lines 19 through 22 insert element $e_i$ between two elements $e_j$ and $e_k$ such that $(e_j \rightarrow next) = e_i = (e_k \rightarrow prev)$. After the insertion, the following is true:

- $QltSlot(e_j) < QltSlot(e_i)$, in which case the invariant is preserved for elements $e_i$ and $e_j$.

- $QltSlot(e_k) > QltSlot(e_i)$, and hence the invariant is also preserved for elements $e_i$ and $e_k$.

Assume now that we invoke the EXTRACT routine on element $e_i$, and let $e_j$ and $e_k$ be the elements before and after $e_i$ in the queue, respectively. There are three possible cases:

- Case 1: $qlt[QltSlot(e_i)] \neq e_i$ (lines 35 and 36). In this case we have that $QltSlot(e_j) \leq QltSlot(e_i)$ and $QltSlot(e_k) = QltSlot(e_i)$. Hence after extracting $e_i$ we have $QltSlot(e_j) \leq QltSlot(e_k)$ and the invariant is preserved.

- Case 2: $qlt[QltSlot(e_i)] = e_i$ and $QltSlot(e_i) = QltSlot(e_j)$ (line 37 and 38). In this case we have that $QltSlot(e_k) > QltSlot(e_i)$, hence $QltSlot(e_j) < QltSlot(e_k)$ and the invariant is preserved.

- Case 3: $qlt[QltSlot(e_i)] = e_i$ and $QltSlot(e_i) \neq QltSlot(e_j)$ (line 39 and 40). In this case we have that $QltSlot(e_j) < QltSlot(e_i)$ and $QltSlot(e_k) > QltSlot(e_i)$. Hence after extracting $e_i$ we have $QltSlot(e_j) < QltSlot(e_k)$ and the invariant is preserved.

§

Let us now state the complexity of the MR-PQ algorithm:

*Lemma 2. Complexity of the MR-PQ algorithm.* The worst case complexity of the MR-PQ algorithm for the INSERT routine is $O(r)$, where $r = (p_{max} - p_{min})/p_\Delta$ is the number of resolution groups supported by the queue. The complexity of the INSERT routine becomes $O(1)$ if there is at least one element in each slot. The complexity of the PEEK, EXTRACTMIN and EXTRACT routines is $O(1)$.

*Proof.* The cost of the INSERT routine is given by the execution of lines 11 and 12, which in the worst case require a linear scan of the *qlt* table, leading to a cost of $O(r)$, where $r = (p_{max} - p_{min})/p_\Delta$ is the size of the table or number of resolution groups. The performance of this method becomes $O(1)$ when the *qlt* table is fully utilized, since the *while* loop in line 11 returns immediately after the first iteration. The rest of the methods (PEEK, EXTRACTMIN and EXTRACT) have a cost of $O(1)$ since they simply require looking up the top element in a doubly-linked list or removing an element from it.

§

A key observation from Lemma 2 is that MR-PQ performs best when the number of elements in the table is very large, because then most of the slots in the *qlt* table will tend to have at least one element and the cost of running the INSERT routine will tend to $O(1)$. This characteristic of MR-PQ queues is beneficial in that it is precisely when the number of elements is very large that one desires to achieve the greatest performance.

In the next two sections we will introduce two variants of the base MR-PQ algorithm which will exploit this property to beat the performance of a standard priority queue. First, we extend the algorithm to support sliding priorities, enabling support for event-driven applications. For this type of applications, we will show that the lower bound cost of $O(1)$ to insert an element can be achieved. Second, we will incorporate the idea of using a binary heap to order the resolution groups to achieve a new worst case lower bound performance in inserting an element of $O(log(r))$. For those applications in which $r$ (the number of resolution groups) is much smaller than $n$ (the number of elements), the proposed algorithm can beat traditional high performance priority queues as we will show.

*3.4. Generalization to Sliding Priorities*

The base MR-PQ algorithm assumes that the priorities from any of the elements in the queue belong to the set $[p_{min}, p_{max})$. In this section, we generalize the base algorithm to support a more flexible schema by allowing the priority set to slide by following the expression $[p_{min} + \delta(t), p_{max} + \delta(t))$, where $\delta(t)$ is a monotonically increasing function of a parameter $t$. The case of sliding priority sets is particularly relevant to applications that run on *event-driven simulators*. (See for example Section 6.5 of [2].) These applications are traditionally implemented using priority queues in which the elements are events to be processed at a future time and their priority field is the time at which these events need to be triggered. In particular, for these applications we have $\delta(t) = t$, where $t$ is time itself. Because events with a lower time value need to be processed before those with a higher value, event-driven simulators are commonly implemented using a *min-priority queues.*

In order to extend the base MR-PQ algorithm to support sliding priorities, we make the following observation applicable to event-driven simulators: if the next event to process has a priority field equal to $p^*$ (corresponding to the time at which the event needs to be processed), then it must be that the application's current time is at least $p^*$ and therefore we can conclude that no other event will be inserted into the queue with a priority value $p^* - p_\Delta$ or lower. As a result, we can set the sentinel's priority field to $p^* - p_\Delta$ and shift the values of the *qlt* table so that the first slot points to the priority resolution group of $p^*$. We implement this shifting feature in a new routine called QLTSLIDE, which is passed a parameter prio corresponding to the priority of the last element removed from the queue $p^*$:

**QLTSLIDE(prio)**

```
44   shift = (prio - queue->prio)/pdelta - 1;
45   if shift < 1:
46       return;
47   for i in [1, qltsize):
48       if i < qltsize - shift:
49           qlt[i] = qlt[i + shift];
50       else
51           qlt[i] = 0;
52   queue->prio = prio - pdelta;
```

Specifically, in line 44 we compute the shift in the priority space that results from extracting a top element with priority prio. In lines 45 and 46, if the shift is less than one slot, there is no need to slide the priority space and the routine returns. (In event-driven simulators, this corresponds to the case in which time has not advanced enough to shift the *qlt* table by one or more slots.) Otherwise, we shift the *qlt* table to the left by shift slots and pad the last shift slots with zeros (lines 47 through 51). In line 52, we update the priority of the sentinel to reflect the shift. (In event-driven simulators, this is equivalent to setting the priority of the sentinel to the current simulator's time.)

We invoke the function QLTSLIDE every time we extract an element from the top of the queue, which requires us to update the routine EXTRACTMIN as follows:

**EXTRACTMIN()**

```
53   e = PEEK();
54   EXTRACT(e);
55   QLTSLIDE(e->prio);    // Added to support sliding
                           // priorities
56   return e;
```

*Lemma 3. Correctness of the MR-PQ algorithm with sliding priorities.* The modified EXTRACTMIN routine preserves the MR-PQ invariant (Property 1).

*Proof.* The only variant introduced consists in invoking the QLTSLIDE method in line 55. Notice that by construction, this method removes slots from the left of the *qlt* table that will no longer be populated (since the priority space has shifted by shift slots) and sets to empty the same amount of slots to the right of the *qlt* table (since those slots have not been populated yet). The remaining slots are not altered which guarantees the preservation of the MR-PQ invariant.

§

In Section 4 we show how the MR-PQ algorithm with sliding priorities can be used to improve the performance of an HPC application.

### 3.5. Binary Heap Based Multiresolution Priority Queues

Lemma 2 shows that the worst case performance to insert an element in the MR-PQ is $O(r)$, where $r$ is the number of resolution groups supported by the queue. This linear cost comes from lines 11 and 12 in the INSERT routine which, in the worst case, require scanning the complete *qlt* table to find an empty slot. We observed that for HPC applications, this cost will be amortized by the fact that the number of elements in the queue will be very high, and so the actual complexity will tend to be $O(1)$ as on average all the slots become filled in with at least one element. (See Section 4 for an actual HPC application operating at this bound.)

For the case in which not all of the slots in the *qlt* table are filled in, we can easily further improve the worst case cost of MR-PQ by using a *binary search tree* to implement the *qlt* table, instead of a vector as is done in the base implementation. Fig. 2 shows the example presented in Fig. 1 using the binary tree optimization.

The result of using a binary search tree is that the cost of searching for the next empty slot in lines 11 and 12 of the INSERT routine can be reduced from $O(r)$ to $O(log(r))$. We will refer to this variant as a *binary tree based multiresolution priority queue* or BT-MR-PQ.

In the next table we summarize the computational cost of the three approaches: the standard binary heap based priority queue (BH-PQ), the base multiresolution priority queue (MR-PQ), and the binary tree based multiresolution priority queue (BT-MR-PQ).

*Table 1. Computational cost.*

| Algorithm | INSERT | PEEK | EXTRACTMIN | EXTRACT |
|---|---|---|---|---|
| BH-PQ | $log(n)$ | $O(1)$ | $log(n)$ | $log(n)$ |
| MR-PQ | $O(r)$ or $O(1)$ | $O(1)$ | $O(1)$ | $O(1)$ |
| BT-MR-PQ | $O(log(r))$ | $O(1)$ | $O(1)$ | $O(1)$ |

To illustrate the benefits of using BT-MR-PQ, consider the problem of finding the shortest path between two nodes in a graph using Dijkstra's algorithm. The standard implementation of Dijkstra relies on using a min-priority queue implemented using a binary heap. To resolve a graph with $v$ vertices and $e$ edges, where each edge has a weight corresponding to its cost, Dijkstra's algorithm requires $v$ INSERT operations, $v$ EXTRACTMIN, and $e$ EXTRACTMIN and INSERT operations. Since in a priority queue implemented as a binary heap the cost of INSERT and EXTRACTMIN is $O(log(v))$, we have that the cost of executing Dijkstra's algorithm is $O((v+e) \cdot log(v))$. (Notice that this bound can be reduced to $O(e + v \cdot log(v))$ if we use a Fibonacci heap [4], but the same general argument applies.) If instead we use a multiresolution priority queue, then the cost becomes $O((v+e) \cdot log(r))$, where $r$ is the number of resolutions in the space of edge weights. Hence for the problems in which $r$ is much smaller than $v$, BT-MR-PQ provides a better bound. Further, if $v$ is very large, from Lemma 2, the MR-PQ base implementation can lead to an amortized cost of the INSERT operation of $O(1)$, making the amortized cost of Dijkstra's algorithm $O(v+e)$. We leave for a future paper the study of the implications of this result to applications in the field of graph theory.

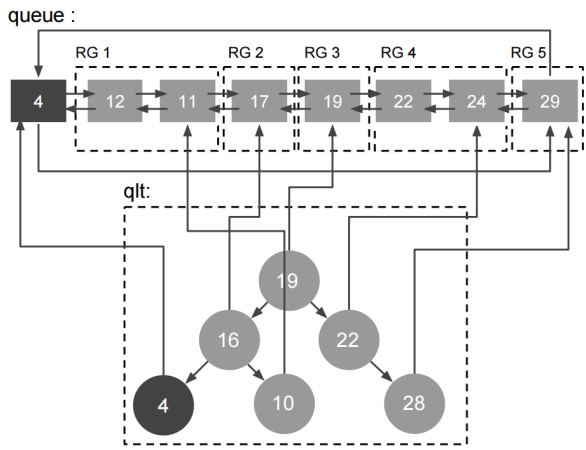

*Fig. 2. The example of MR-PQ in Fig. 1, using a binary tree to implement the qlt table.*

## 4. IMPLEMENTATION AND BENCHMARKS

We have implemented and tested multiresolution priority queues to improve the performance of Bro, an event-driven application which enables deep network analytics [12]. Bro uses a standard priority queue implemented as a binary heap to manage the system's timers. While binary heaps are excellent implementations of priority queues, we find that when dealing with very high speed traffic (e.g., at rates of 10Gbps), they become a system bottleneck. This is illustrated in Fig. 3, which shows a call graph profile of the computational cost (measured in percentage of CPU usage) incurred by some of the main functional parts of the Bro stack. Each box presents a profiled function and includes two percentage values: the first reflects the cost of that function divided by the total aggregated computational cost of Bro; the second percentage value reflects the cost of that function and all of its children functions divided by the total aggregated cost. The next list illustrates the top functions in Bro ordered by their computational cost:

```
Total: 63724 samples
    4139    6.5% PriorityQueue::BubbleDown
    2500    3.9% SLL_Pop
    1899    3.0% Ref
    1829    2.9% Unref
    1701    2.7% PackedCache::KeyMatch
    1537    2.4% Attributes::FindAttr
    1249    2.0% Dictionary::Lookup
    1184    1.9% NameExpr::Eval
```

As shown, the function `PriorityQueue::BubbleDown` takes the top spot with a cost of 6.5% of the total processing cost. This function implements the standard *bubble down* operation (sometimes also know as *heapify* [2]) part of the binary heap based implementation of a priority queue. Specifically, Bro uses this method every time it needs to fetch the next element from the queue of timers.

While binary heaps provide an efficient implementation of priority queues—with the `BubbleDown` function having a computational cost of $O(log(n))$, where $n$ is the number of elements in the queue—the above profiling results illustrate that they can still become a critical bottleneck in real world HPC applications. This is especially relevant if we consider that Bro's implementation of a priority queue corresponds to 105 lines of C++ code, whereas the project includes a total of 333,930 lines of C/C++ code (without accounting for the many lines of non-C interpreted code run by the engine). In other words, the priority queue implementation corresponds to 0.03% of the total C/C++ code in Bro, but it takes a computational cost of 6.5%.

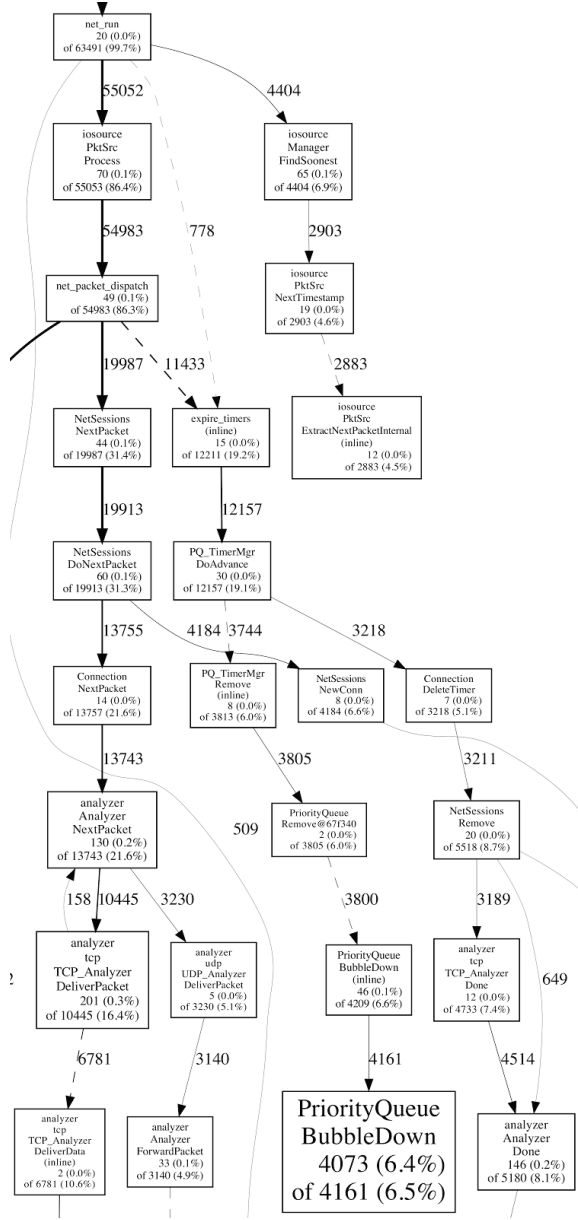

*Fig. 3. Binary heap based priority queues can become a system bottleneck.*

This result is explained by the sheer number of timers Bro's priority queue needs to handle. At rates of 10Gbps, it may need to handle in the order of tens of thousands of connections originated per second. In Bro, each connection sets multiple timers to manage its various protocol states (examples of timers in Bro include connection establishment, connection inactivity or connection linger timeouts, among others), potentially leading to hundreds of thousands of timers generated every second. Under this scenario, we observe that:

- Bro's timeout values are represented as *double* type, effectively enabling a priority set with infinite resolution in the space of real numbers $\mathbb{R}$.
- However, all timeout values in Bro are at least 1 second or more.

The above presents a mismatch between the application requirements and the capabilities provided by the implementation. The implementation unnecessarily provides infinite resolution timers, which leads to unnecessary processing overhead. Multiresolution priority queues allow applications to tune the priority resolution of the implementation to match the necessary requirements (no more and no less) to achieve better performance. In our application, choosing an MR-PQ with parameter $p_\Delta = 1$ second is enough, since no Bro timer requires a resolution lower than 1 second. Further, notice that in processing network traffic, more than 1 timer is generated for each 1 second resolution group (indeed, in the order of tens to hundreds of thousands of timers are generated every second at 10Gbps), which means that the performance of MR-PQ will be optimal at $O(1)$ (see Lemma 2).

We have modified Bro's priority queue with the sliding version of the MR-PQ algorithm described in Sections 3.2 and 3.4. The resulting implementation takes a total of 144 lines of C++ code.

We run Bro as a network traffic analysis engine on a host equipped with two Intel Xeon E5-2670 processors clocked at 2.50 GHz for a total of 20 physical cores with 25.6MB of L3 cache for each processor and 4 x 10 Gbps SFP optical network interfaces. Traffic is load balanced onto a cluster of Bro nodes. (See [13] for a description of Bro's cluster architecture.)

The system is fed with a traffic dataset generated from a mix of real packet traces from our corporate network in New York and synthetically generated traffic using an Ixia traffic generator, resulting in a dataset of human generated traffic (for applications such as HTTP/HTTPS) and machine generated flows (for services such as SNMP). Table 2 broadly summarizes the traffic dataset.

*Table 2. Statistics of the traffic dataset.*

| TCP | UDP | ICMP | Other | Avg pkt size (B) | Avg conn size (KB) |
|---|---|---|---|---|---|
| 92.34% | 7.5% | 0.02% | 0.04 | 510.25 | 7050.16 |

Fig. 4 through Fig. 8 present various performance metrics resulting from running Bro with its default binary heap priority queue (BH-PQ) and with our implementation of the multiresolution priority queue (MR-PQ) against our traffic dataset at various rates from 2 Gbps up to 10 Gbps.

Fig. 4 and Fig. 5 focus on finer grained CPU level metrics by providing the performance of the cache and the CPU instruction pipeline. We observe that MR-PQ improves both the cache performance and the instructions per cycle. Improvements in the cache performance can be explained from the reduction of the cost of inserting an element into the priority queue from $O(log(n))$ for a standard priority queue to $O(1)$ for MR-PQ. For each incoming connection, Bro's `BubbleDown` function requires scanning multiple times (one for each timer type) the timers' binary heap, which leads to excessive cache thrashing. On the other hand, MR-PQ simply requires linking a single element into a doubly linked list.

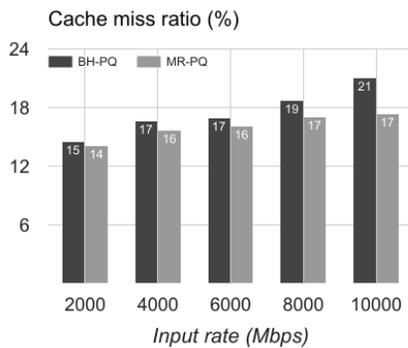

Fig. 4. Cache miss ratio.

In Fig. 5, the reduction in the instructions per cycle (IPC) can be explained as a consequence of the improvements on the cache performance. Since MR-PQ reduces cache miss ratios, this leads to lower stalling cycles waiting for data fetches from main memory to reload the cache, which leads to better scheduling of the CPU instruction pipeline and hence an improved IPC metric.

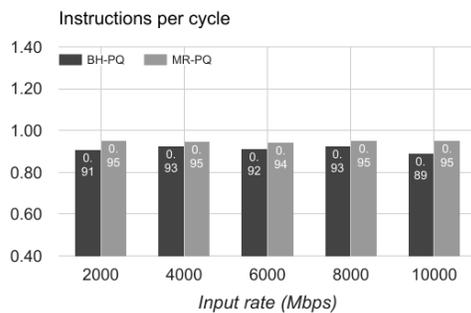

Fig. 5. Instruction per cycle (IPC).

Fig. 6, 7 and 8 focus on system wide performance metrics. Fig. 6 provides total percentage of packets dropped by the network interface card (NIC). These packets are dropped by the NIC due to the Bro application not being able to process packets fast enough. Again thanks to removing the bottleneck produced by the `BubbleDown` function, MR-PQ substantially reduces packet drops across the board for all traffic rates.

In a highly complex system like Bro which includes many layers of packet processing (network, protocol analyzers, high level analytics, logging of metadata into disk, housekeeping, etc.), it is often difficult to optimize the system for one parameter and to obtain improvements for all performance metrics. A most common error in performing system wide optimization is to design the system for maximum packet throughput. In such cases, often a reduction in packet drops lead to penalties in the amount of output metadata generated by the system, because by forcing the system to ingest more packets than it can possibly process, non-linear negative effects (such as an increase in cache thrashing) lead to lower system wide performance [14]. To verify that our optimization does not only reduce packet drops but also improve system wide performance, we measure the total number of events generated by Bro. (Events in Bro are responsible to generate the metadata extracted from the traffic and hence can be understood as a system wide performance metric.) As shown in Fig. 7, with MR-PQ the system generates also more events. To reinforce this point, in Fig. 8 we also present the amount of files carried by network flows (from protocols such as SMTP, FTP, etc.) that Bro can process, showing also an improvement when using MR-PQ.

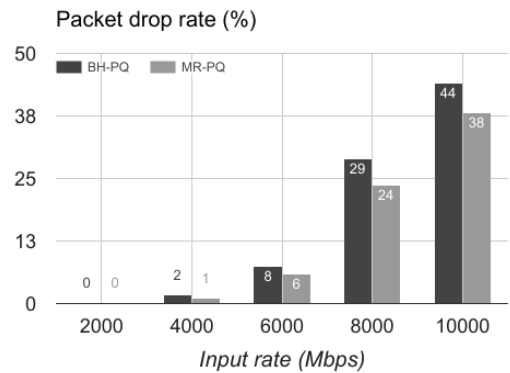

Fig. 6. Packet drop rate.

In summary, the implementation of Bro's timer manager using a multiresolution priority queue (a change of 144 lines of C++ code) leads to performance improvements of both fine grained CPU metrics and across all measured system wide performance metrics thanks to reducing the main cost of inserting an element into the priority queue from $O(log(n))$ to amortized $O(1)$.

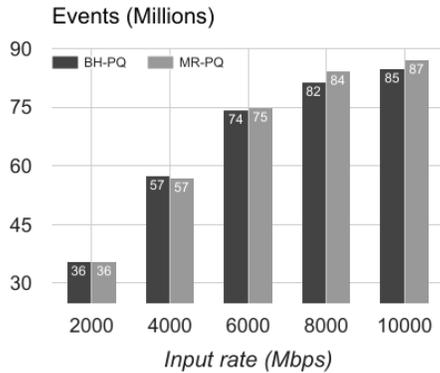

*Fig. 7. Total network events generated.*

5. CONCLUSIONS

The implementation of high performance priority queues has been the subject of intense research since the binary heap data structure and the heapsort algorithm were invented in 1964. To the best of our knowledge, all existing work has focused on solving the total ordering problem, whereby for any pair of elements $e_i$ and $e_j$, either $e_i \leq_p e_j$ or $e_j \leq_p e_i$, for some priority ordering $\leq_p$. In this paper we propose to relax this requirement to design a priority queue that only guarantees the partial ordering of its elements. The result is a new data structure called *multiresolution priority queue* which allows the application to trade off a small resolution error in the set of priorities to achieve better performance.

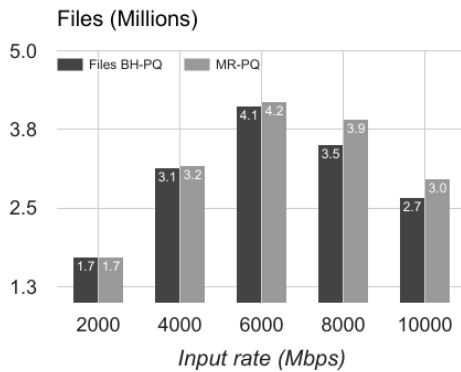

*Fig. 8. Total files processed.*

The new data structure reduces the worst case performance of inserting an element from $O(log(n))$ to $O(log(r))$, where $n$ is the number of elements in the queue and $r$ is the number of resolution groups in the priority space. In many applications $r \ll n$, which makes multiresolution priority queues very efficient data structures. Further, when the number of elements in the table is high, the amortized cost to insert an element becomes $O(1)$.

An interesting result is that if the application deals with elements in a priority set that can be organized into $r$ resolution groups where each group has only one possible priority value, then the better performance gain of a multiresolution priority queue comes free of any error. Forthcoming work will focus on further exploring how this property can be used to accelerate the performance of essential algorithms such as finding the shortest path between two nodes in a graph which are core to so many HPC applications in the field of computer science.